\documentclass[prd,reprint,showpacs,showkeys]{revtex4-1}
\usepackage{amsfonts}
\usepackage{mdframed}
\usepackage{amssymb}
\usepackage{amsmath}
\usepackage{graphicx}
\usepackage[font={footnotesize,it}]{caption}

\begin{document}

\title{2+1-dimensional traversable wormholes supported by positive energy}
\author{S. Habib Mazharimousavi}
\email{habib.mazhari@emu.edu.tr}
\author{M. Halilsoy}
\email{mustafa.halilsoy@emu.edu.tr}
\affiliation{Department of Physics, Eastern Mediterranean University, Gazima\u{g}usa,
north Cyprus, Mersin 10, Turkey. }
\date{\today }

\begin{abstract}
We revisit the shapes of the throats of wormholes, including thin-shell
wormholes (TSWs) in $2+1-$dimensions. In particular, in the case of TSWs
this is done in a flat $2+1-$dimensional bulk spacetime by using the
standard method of cut-and-paste. Upon departing from a pure time-dependent
circular shape i.e., $r=a\left( t\right) $ for the throat, we employ a $%
\theta -$dependent closed loop of the form $r=R\left( t,\theta \right) ,$
and in terms of $R\left( t,\theta \right) $ we find the surface energy
density $\sigma $ on the throat. For the specific convex shapes we find that
the total energy which supports the wormhole is positive and finite. In
addition to that we analyze the general wormhole's throat. By considering a
specific equation of $r=R\left( \theta \right) $ instead of $r=r_{0}=const.,$
and upon certain choices of functions for $R\left( \theta \right) $ we find
the total energy of the wormhole to be positive.
\end{abstract}

\maketitle

\section{Introduction}

In the theory of wormholes the prime important issue concerns energy which
turns out to be negative (i. e. exotic matter) to resist against
gravitational collapse. This and stability related matters informed by
Morris and Thorne \cite{MT} were restructured later on by Hochberg and
Visser \cite{HV}. The 2+1-dimensional version of wormholes was considered
first in \cite{Mann} and more recent works can be found in \cite{Revised}.
Nonexistence of negative energy in classical physics / Einstein's general
relativity persisted as a serious handicap. Given this fact, and without
reference to quantum theory in which negative energy has rooms at smallest
scales to resolve the problem of microscopic wormholes, how can then one
tackle with the large scale wormholes?. In this study, first we restrict
ourselves to thin-shell wormholes (TSWs) which are tailored by the
cut-and-paste technique of spacetimes \cite{MV,MV0}. In our view the
theorems proved in \cite{HV} for general wormholes should be taken
cautiously and mostly relaxed when the subject matter is TSWs \cite{Habib1}.
One important point that we emphasize / exploit is that the throat need not
have a circular topology. It may depend on the angular variable as well, for
example. This is the case that we naturally confront in static,
non-spherical spacetimes. One such example is the Zipoy-Voorhees
(ZV)-geometry which deviates from spherical symmetry by a deformation /
oblateness parameter \cite{ZV}. We employed this to show that the overall /
total energy can be made positive although locally, depending on angular
location it may take negative values \cite{Habib2}. We construct the
simplest possible TSW in $2+1-$dimensions whose bulk is made of flat
Minkowski spacetime. Such a wormhole was constructed first by Visser \cite%
{MV0}, for the spherical throat case in $3+1-$dimensions. In our case of $%
2+1-$dimensions the only non-zero curvature is at the throat which consists
of a ring, apt for the proper junction conditions. For the shape of the
throat we assume an arbitrary angular dependence in order to attain
ultimately a positive total energy (i.e. normal matter). In other words, the
throat surface is chosen as $F=r-R\left( t,\theta \right) =0,$ for an
appropriate function $R\left( t,\theta \right) .$ For $R\left( t,\theta
\right) =a\left( t\right) $, we recover the circular topology considered to
date. Note that $t$ is the coordinate time measured by an external observer.
As a possible choice we employ $R\left( 0,\theta \right) =R_{0}\left( \theta
\right) =\frac{1}{\sqrt{\left\vert \cos \frac{5}{2}\theta \right\vert }+1}$
which represents a starfish shape. The crucial point about the path of the
throat is that it must be convex rather than concave in order to attain
anything but exotic matter. A circle, which is concave yields the undesired
negative energy. We present various alternatives for the starfish
geometrical shapes to justify our argument. The limiting case of almost zero
periodic dependence on the angle brings us to that of total energy zero(=a
vacuum) on the throat, which amounts to making TSW from a vacuum.

With this much information about $2+1-$dimensional TSWs we extend our
argument to the $2+1-$dimensional general traversable wormhole which is
considered to be a brane in $3+1-$dimensional flat spacetime. We provide
explicit examples to show that $2+1-$dimensional wormholes can be fueled by
a total positive energy and the null energy condition is satisfied.

Organization of the paper is as follows. In Section II we study TSWs with
general throat shapes. $2+1-$dimensional wormholes induced from $3+1-$%
dimensional flat spacetime is considered in Section III. The paper ends with
Conclusion in Section IV.

\section{Thin-shell wormholes with general throat shape}

In this section we consider a model of TSW in $2+1-$dimensional flat
spacetime. Hence, the bulk metric is given by%
\begin{equation}
ds^{2}=-dt^{2}+dr^{2}+r^{2}d\theta ^{2}.
\end{equation}%
Following \cite{MV} we introduce $M^{\pm }=\left\{ r>R\left( t,\theta
\right) \right\} $ as two incomplete manifolds from the original bulk and
then we paste them on an identical hypersurface with equation 
\begin{equation}
F\left( t,r,\theta \right) =r-R\left( t,\theta \right) =0
\end{equation}%
to make upon them a complete manifold known as the TSW. Let's note that, in 
\cite{MV} and the other consequent papers the proper time $\tau $ is used
instead of the coordinate time $t$ which we consider here. As one can see,
the results found in terms of $t$ can be easily transformed to terms of $%
\tau $. For instance, in the case we consider here by setting $\left( \frac{%
\partial R}{\partial t}\right) ^{2}=\left( \frac{\partial R}{\partial \tau }%
\right) ^{2}/\left( 1+\left( \frac{\partial R}{\partial \tau }\right)
^{2}\right) $ the results will be expressed in terms of $\tau .$ This
relation also shows that in the static equilibrium case $\frac{\partial R}{%
\partial t}=\frac{\partial R}{\partial \tau }=0$ and therefore the physical
properties such as the energy density and pressures are the same.

The throat is located on the shell $r=R\left( t,\theta \right) $ and
therefore $R\left( t,\theta \right) $ is a general function of $\theta $ and 
$t$ but not arbitrary. As, $r=R\left( t,\theta \right) $ is going to be the
throat which connects two different spacetimes, in $2+1-$dimensions it must
be a closed loop. We choose $x^{\alpha }=\left( t,r,\theta \right) $ for the
bulk and $\xi ^{i}=\left( t,\theta \right) $ for the hypersurface. Therefore
while the bulk metric is given by 
\begin{equation}
g_{\mu \nu }=diag\left( -1,1,r^{2}\right) ,
\end{equation}%
the induced metric on the shell $h_{ij}$ is obtained by using 
\begin{equation}
h_{ij}=\frac{\partial x^{\alpha }}{\partial \xi ^{i}}\frac{\partial x^{\beta
}}{\partial \xi ^{j}}g_{\alpha \beta }.
\end{equation}%
One finds%
\begin{equation}
ds_{\Sigma }^{2}=-\left( 1-\dot{R}^{2}\right) dt^{2}+\left( R^{2}+R^{\prime
2}\right) d\theta ^{2}+2\dot{R}R^{\prime }dtd\theta
\end{equation}%
in which a prime and a dot stand for derivative with respect to $\theta $
and $t,$ respectively. Next, we find the extrinsic curvature tensor defined
as%
\begin{equation}
K_{ij}^{\pm }=-n_{\gamma }^{\pm }\left( \frac{\partial ^{2}x^{\gamma }}{%
\partial \xi ^{i}\partial \xi ^{j}}+\Gamma _{\alpha \beta }^{\gamma }\frac{%
\partial x^{\alpha }}{\partial \xi ^{i}}\frac{\partial x^{\beta }}{\partial
\xi ^{j}}\right) ,
\end{equation}%
in which 
\begin{equation}
n_{\gamma }^{\pm }=\pm \frac{1}{\sqrt{\Delta }}\frac{\partial F\left(
t,r,\theta \right) }{\partial x^{\gamma }}
\end{equation}%
with%
\begin{equation}
\Delta =\frac{\partial F\left( t,r,\theta \right) }{\partial x^{\alpha }}%
\frac{\partial F\left( t,r,\theta \right) }{\partial x^{\beta }}g^{\alpha
\beta }.
\end{equation}%
Using (2), we find%
\begin{equation}
\Delta =1+\left( \frac{R^{\prime }}{R}\right) ^{2}-\dot{R}^{2}.
\end{equation}%
The exact form of the normal vector is found to be%
\begin{equation}
n_{t}^{\pm }=\pm \frac{1}{\sqrt{\Delta }}\left( -\dot{R}\right) ,
\end{equation}%
\begin{equation}
n_{r}^{\pm }=\pm \frac{1}{\sqrt{\Delta }}
\end{equation}%
and%
\begin{equation}
n_{\theta }^{\pm }=\pm \frac{1}{\sqrt{\Delta }}\left( -R^{\prime }\right) ,
\end{equation}%
such that $\left\vert \vec{n}\right\vert =1.$ The bulk's line element (1)
admits $\Gamma _{r\theta }^{\theta }=\Gamma _{\theta r}^{\theta }=\frac{1}{r}%
,$ $\Gamma _{\theta \theta }^{r}=-r$ while the rest of Christoffel symbols
are zero. Therefore, the extrinsic curvature tensor elements become%
\begin{equation}
K_{t\theta }^{\pm }=-n_{r}^{\pm }\dot{R}^{\prime }-n_{\theta }^{\pm }\left( 
\frac{\dot{R}}{R}\right)
\end{equation}%
\begin{equation}
K_{tt}^{\pm }=-n_{r}^{\pm }\ddot{R}
\end{equation}%
\begin{equation}
K_{\theta \theta }^{\pm }=-n_{r}^{\pm }\left( R^{\prime \prime }-R\right)
-2n_{\theta }^{\pm }\frac{R^{\prime }}{R}.
\end{equation}%
Israel junction conditions \cite{Israel} read%
\begin{equation}
k_{i}^{j}-k\delta _{i}^{j}=-8\pi S_{i}^{j}
\end{equation}%
in which 
\begin{equation}
S_{i}^{j}=\left( 
\begin{array}{cc}
-\sigma & q_{1} \\ 
q_{2} & p%
\end{array}%
\right)
\end{equation}%
is the energy-momentum tensor on the thin-shell, $%
k_{i}^{j}=K_{i}^{j+}-K_{i}^{j-}$ and $k=trace\left( k_{i}^{j}\right) .$ Note
that $q_{1}$ and $q_{2}$ are appropriate pressure terms. Combining the
results found above we get%
\begin{equation}
k_{tt}=-\frac{2\ddot{R}}{\sqrt{\Delta }}
\end{equation}%
\begin{equation}
k_{\theta \theta }=\frac{-2}{\sqrt{\Delta }}\left( R^{\prime \prime }-R-%
\frac{2R^{\prime 2}}{R}\right)
\end{equation}%
and%
\begin{equation}
k_{t\theta }=\frac{-2}{\sqrt{\Delta }}\left( \dot{R}^{\prime }-\frac{%
R^{\prime }\dot{R}}{R}\right) .
\end{equation}%
Furthermore, one finds%
\begin{equation}
k_{t}^{t}=h^{tt}k_{tt}+h^{t\theta }k_{t\theta }
\end{equation}%
\begin{equation}
k_{\theta }^{\theta }=h^{\theta \theta }k_{\theta \theta }+h^{\theta
t}k_{\theta t}
\end{equation}%
\begin{equation}
k_{t}^{\theta }=h^{\theta t}k_{tt}+h^{\theta \theta }k_{t\theta }
\end{equation}%
and%
\begin{equation}
k_{\theta }^{t}=h^{tt}k_{\theta t}+h^{t\theta }k_{\theta \theta }.
\end{equation}%
We recall that%
\begin{equation}
h_{ij}=\left( 
\begin{array}{cc}
-B & H \\ 
H & A%
\end{array}%
\right)
\end{equation}%
which implies%
\begin{equation}
h^{ij}=\left( 
\begin{array}{cc}
\frac{-A}{AB+H^{2}} & \frac{H}{AB+H^{2}} \\ 
\frac{H}{AB+H^{2}} & \frac{B}{AB+H^{2}}%
\end{array}%
\right)
\end{equation}%
in which $A=\left( R^{\prime 2}+R^{2}\right) $, $B=1-\dot{R}^{2}$ and $H=%
\dot{R}R^{\prime }.$ Considering (26) in (21-25) we find%
\begin{equation}
k_{t}^{t}=\frac{-2\left[ \left( 1-\dot{R}^{2}\right) \left( \dot{R}^{\prime
}-\frac{R^{\prime }\dot{R}}{R}\right) -\left( R^{\prime 2}+R^{2}\right) 
\ddot{R}\right] }{\left[ \left( R^{\prime 2}+R^{2}\right) \left( 1-\dot{R}%
^{2}\right) +\dot{R}^{2}R^{\prime 2}\right] \sqrt{1+\left( \frac{R^{\prime }%
}{R}\right) ^{2}-\dot{R}^{2}}}
\end{equation}%
\begin{equation}
k_{\theta }^{\theta }=\frac{-2\left[ \left( 1-\dot{R}^{2}\right) \left(
R^{\prime \prime }-R-\frac{2R^{\prime 2}}{R}\right) +\dot{R}R^{\prime
}\left( \dot{R}^{\prime }-\frac{R^{\prime }\dot{R}}{R}\right) \right] }{%
\left[ \left( R^{\prime 2}+R^{2}\right) \left( 1-\dot{R}^{2}\right) +\dot{R}%
^{2}R^{\prime 2}\right] \sqrt{1+\left( \frac{R^{\prime }}{R}\right) ^{2}-%
\dot{R}^{2}}}
\end{equation}%
\begin{equation}
k_{t}^{\theta }=\frac{-2\left[ \dot{R}R^{\prime }\ddot{R}+\left( 1-\dot{R}%
^{2}\right) \left( \dot{R}^{\prime }-\frac{R^{\prime }\dot{R}}{R}\right) %
\right] }{\left[ \left( R^{\prime 2}+R^{2}\right) \left( 1-\dot{R}%
^{2}\right) +\dot{R}^{2}R^{\prime 2}\right] \sqrt{1+\left( \frac{R^{\prime }%
}{R}\right) ^{2}-\dot{R}^{2}}}
\end{equation}%
and%
\begin{equation}
k_{\theta }^{t}=\frac{-2\left[ \dot{R}R^{\prime }\left( R^{\prime \prime }-R-%
\frac{2R^{\prime 2}}{R}\right) -\left( R^{\prime 2}+R^{2}\right) \left( \dot{%
R}^{\prime }-\frac{R^{\prime }\dot{R}}{R}\right) \right] }{\left[ \left(
R^{\prime 2}+R^{2}\right) \left( 1-\dot{R}^{2}\right) +\dot{R}^{2}R^{\prime
2}\right] \sqrt{1+\left( \frac{R^{\prime }}{R}\right) ^{2}-\dot{R}^{2}}}.
\end{equation}%
Therefore the Israel junction conditions imply%
\begin{equation}
\sigma =-\frac{1}{8\pi }k_{\theta }^{\theta }
\end{equation}%
and%
\begin{equation}
p=\frac{1}{8\pi }k_{t}^{t}.
\end{equation}%
In static equilibrium, one may set $R=R_{0}\left( \theta \right) $ and $\dot{%
R}_{0}=\ddot{R}_{0}=0$ which consequently yield%
\begin{equation}
\sigma _{0}=\frac{1}{4\pi }\frac{\left( R_{0}^{\prime \prime }-R_{0}-\frac{%
2R_{0}^{\prime 2}}{R_{0}}\right) }{\left( R_{0}^{\prime 2}+R_{0}^{2}\right) 
\sqrt{1+\left( \frac{R_{0}^{\prime }}{R_{0}}\right) ^{2}}},
\end{equation}%
and%
\begin{equation}
p_{0}=q_{10}=q_{20}=0.
\end{equation}%
This is not surprising since the bulk spacetime is flat. Therefore in the
static equilibrium, the only nonzero component of the energy-momentum tensor
on the throat is the energy density $\sigma _{0}.$ We note that the total
matter supporting the wormhole is given by%
\begin{equation}
\Omega =\int_{0}^{2\pi }\int_{0}^{\infty }\sqrt{-g}\sigma \delta \left(
r-R\right) drd\theta .
\end{equation}%
Let's add that, a physically acceptable energy-momentum tensor must at least
satisfy the weak energy conditions which implies i) $\sigma _{0}\geq 0$ and
ii) $\sigma _{0}+p_{0}\geq 0$ and the total energy to be positive and
finite. Therefore for the energy momentum tensor which we have found on the
surface with only energy density nonzero, our task reduces naturally to find
the specific case with $\sigma _{0}\geq 0$ and $0\leq \Omega <\infty .$

In the sequel we consider the various possibilities of the shape of the
throat including the circular one. The first case to be checked is the
circular throat i.e., $R_{0}=$constant. This leads to $\sigma _{0}=\frac{-1}{%
4\pi }\frac{1}{R_{0}}$ and clearly violates the null energy condition which
states that $\sigma _{0}+p_{0}\geq 0.$ 
\begin{figure}[h]
\includegraphics[width=90mm,scale=0.7]{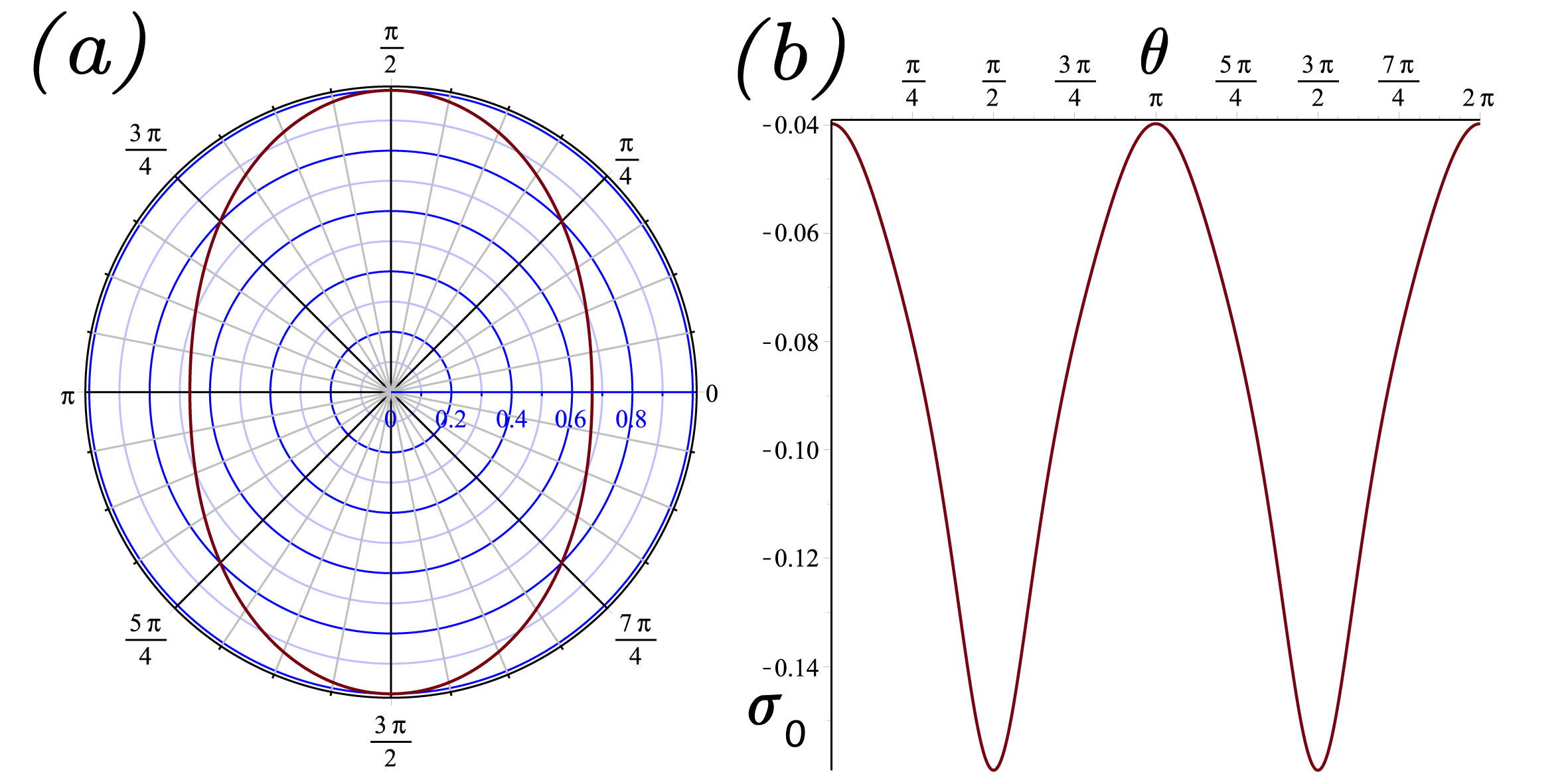}
\caption{The geometry of the throat for $R_{0}\left( \protect\theta \right) =%
\frac{1}{0.5\cos ^{2}\protect\theta +1}$ with its energy density
distribution $\protect\sigma _{0}$. We see that the signature of the
curvature is positive everywhere and as a result the matter is exotic
everywhere.}
\end{figure}
\begin{figure}[h]
\includegraphics[width=90mm,scale=0.7]{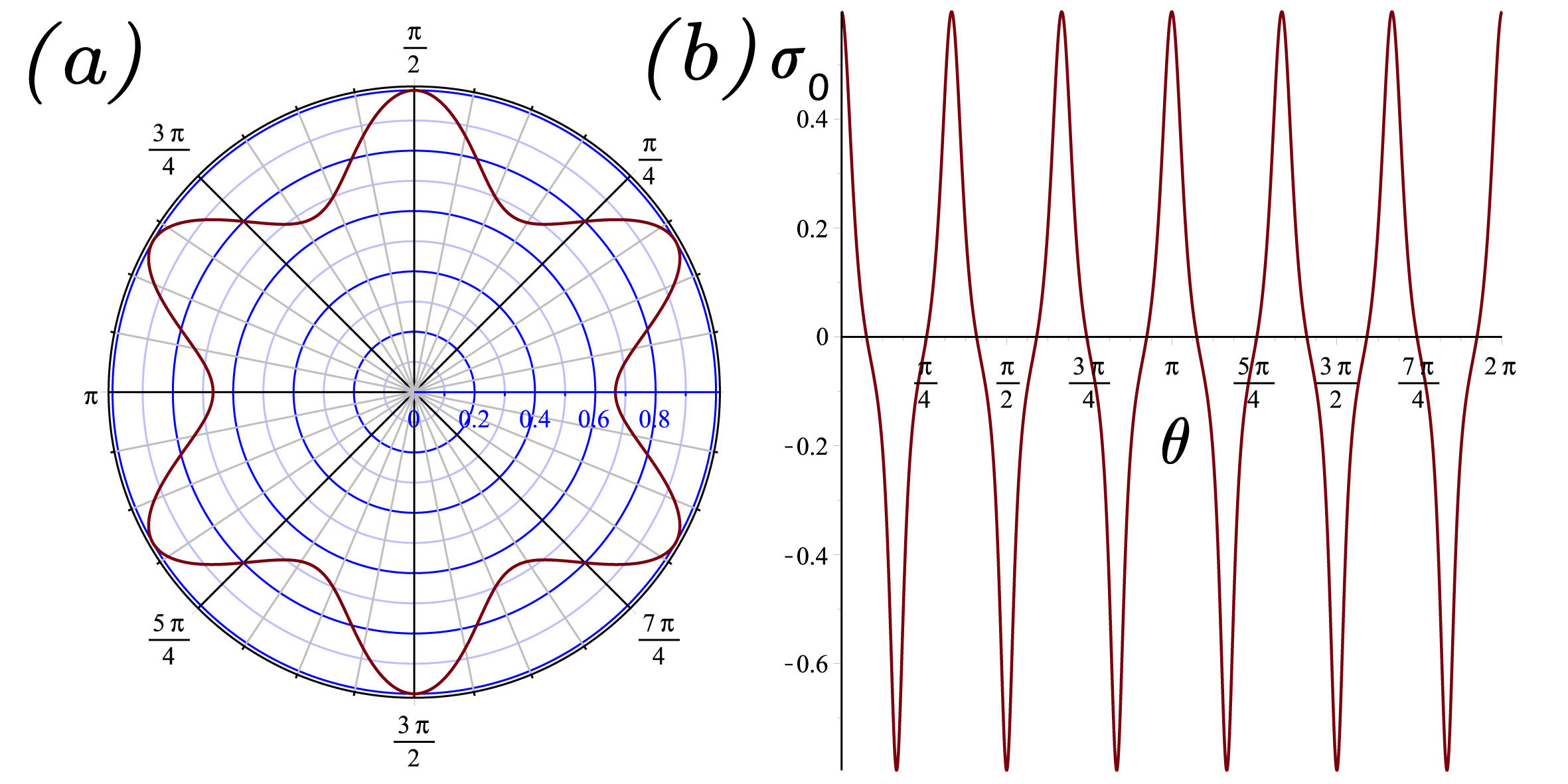}
\caption{The geometry of the throat when $R_{0}\left( \protect\theta \right)
=\frac{1}{0.5\cos ^{2}\left( 3\protect\theta \right) +1}$ and its energy
density distribution $\protect\sigma _{0}$. This figure shows that $\protect%
\sigma _{0}$ is positive when the curvature is negative and vice versa. The
total energy which supports the wormhole, however, is negative.}
\end{figure}
\begin{figure}[h]
\includegraphics[width=90mm,scale=0.7]{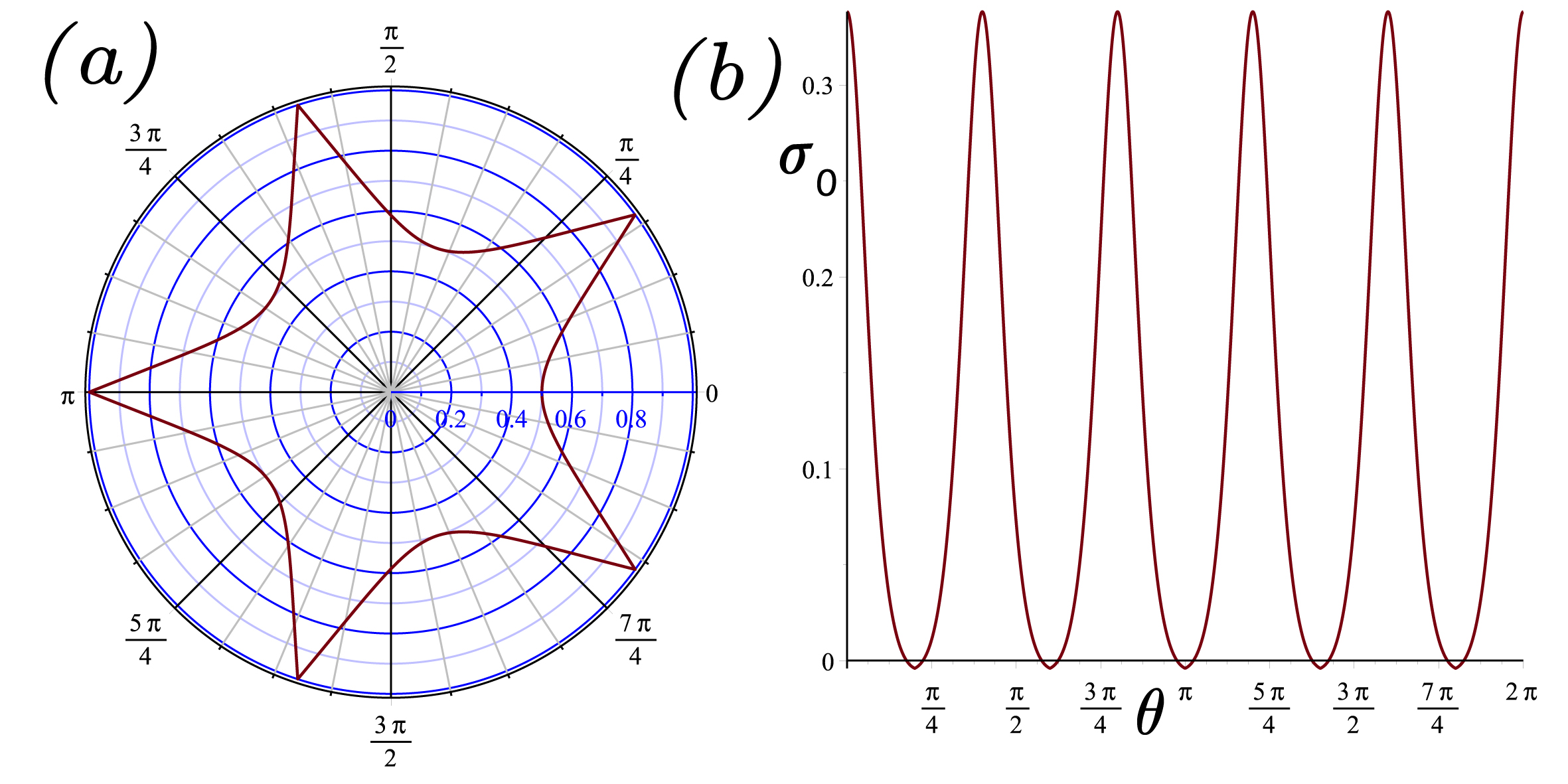}
\caption{The geometry of the throat when $R_{0}\left( \protect\theta \right)
=\frac{1}{\left\vert \cos \left( \frac{5}{2}\protect\theta \right)
\right\vert +1}$ and $\protect\sigma _{0}$ in terms of $\protect\theta .$
Note that $\protect\sigma _{0}$ is overwhelmingly positive and so is the
total energy.}
\end{figure}
\begin{figure}[h]
\includegraphics[width=90mm,scale=0.7]{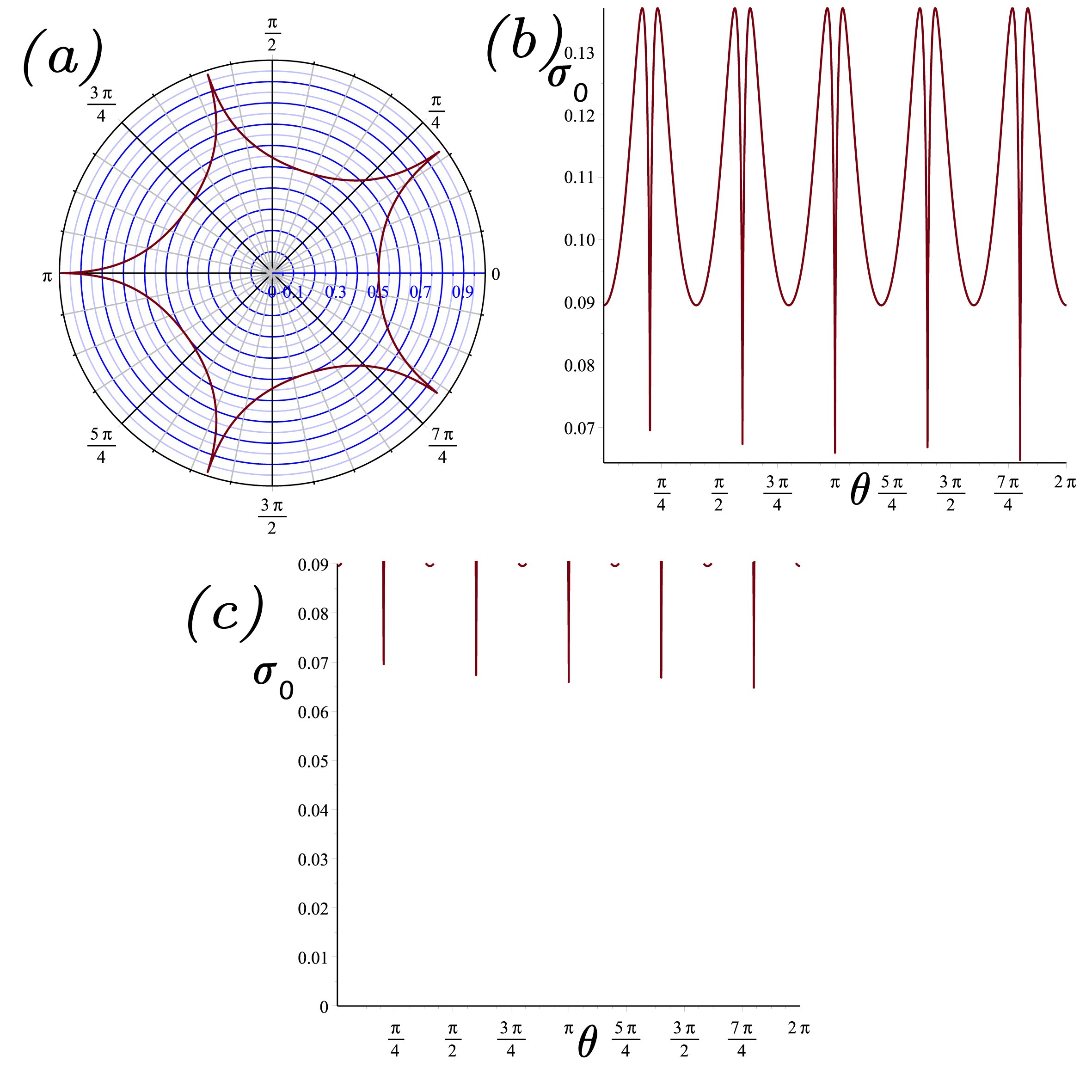}
\caption{The geometry of the throat when $R_{0}\left( \protect\theta \right)
=\frac{1}{\protect\sqrt{\left\vert \cos \left( \frac{5}{2}\protect\theta %
\right) \right\vert }+1}$ and $\protect\sigma _{0}$ in terms of $\protect%
\theta .$ In this figure $\protect\sigma _{0}$ is positive everywhere as the
curvature is negative, and the total energy is positive.}
\end{figure}
\begin{figure}[h]
\includegraphics[width=90mm,scale=0.7]{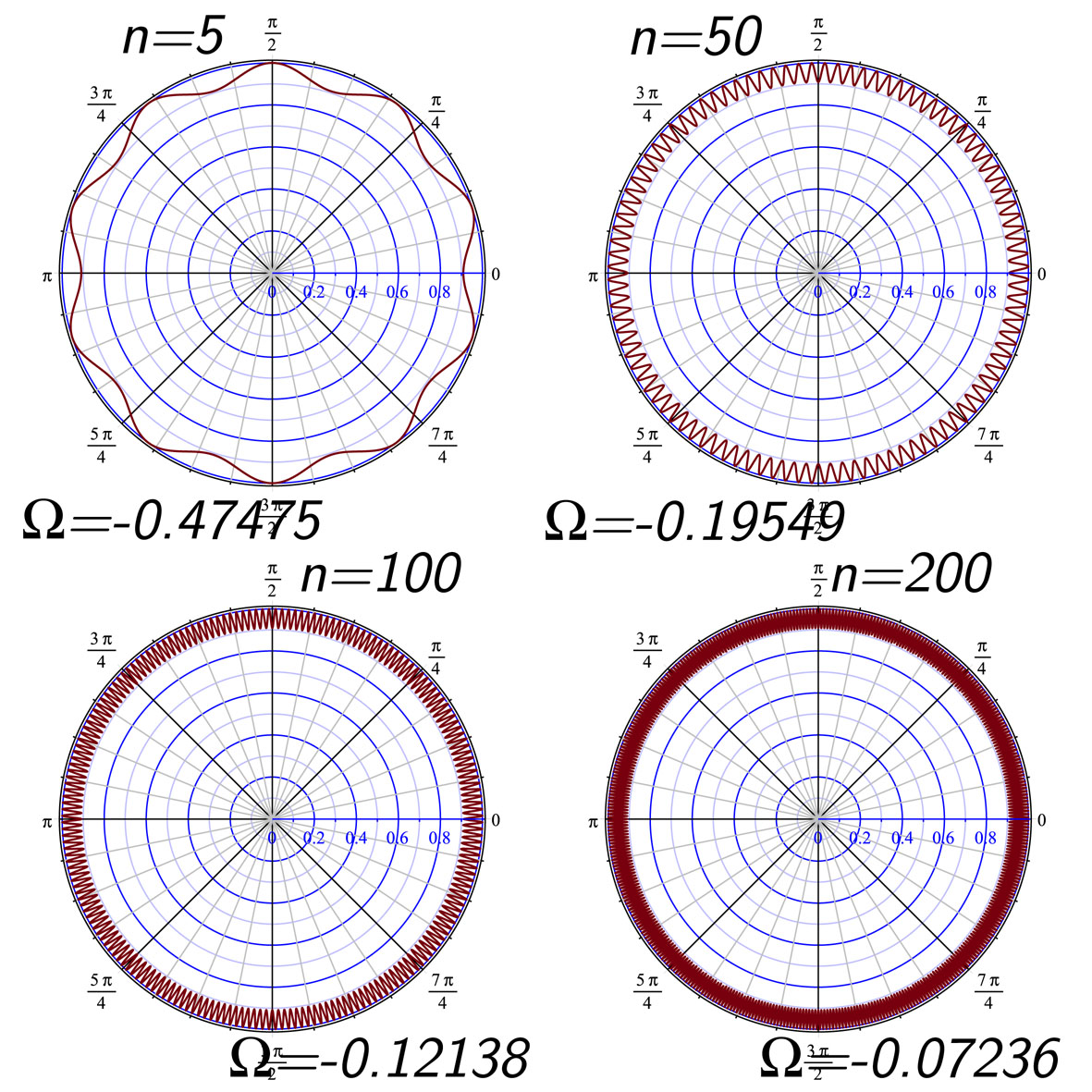}
\caption{The geometry of the throat when $R_{0}\left( \protect\theta \right)
=\frac{1}{\protect\epsilon \cos ^{2}\left( n\protect\theta \right) +1}$ in
terms of $\protect\theta $ for $\protect\epsilon =0.1$ and $n=5,50,100$ and $%
200.$ The total exotic matter $\Omega $ for each case is also given. We
observe that larger $n$ yields smaller amount of exotic matter such that $%
\lim_{n\rightarrow \infty }\Omega =0$.}
\end{figure}

For a specific function of $R_{0}\left( \theta \right) ,$ there are four
different possibilities: i) $\sigma _{0}<0$ on entire domain of $\theta \in %
\left[ 0,2\pi \right] ,$ ii) $\sigma _{0}\leq 0$ or $\sigma _{0}\geq 0$ but
the total energy $\Omega <0,$ iii) $\sigma _{0}\leq 0$ or $\sigma _{0}\geq 0$
with the total energy $\Omega >0$ and iv) $\sigma \geq 0.$ Herein, 
\begin{equation}
\Omega =\int_{0}^{2\pi }R_{0}\sigma _{0}d\theta ,
\end{equation}%
is the total energy on the throat. In what follows we present illustrative
examples for all cases. We note that the specific cases given below can be
easily replaced by other functions but we must keep in mind that although $%
R_{0}\left( \theta \right) $ is a general function, $r=R_{0}$ must present a
closed path in $2+1-$dimensions.

\subsection{ $\protect\sigma _{0}<0,\Omega <0$}

In the first example, the throat is deformed from a perfect circle to an
oval shape given by%
\begin{equation}
R_{0}\left( \theta \right) =\frac{1}{0.5\cos ^{2}\theta +1}.
\end{equation}%
The shape of the throat and $\sigma _{0}$ are shown in Fig. 1a and 1b,
respectively. As we observe here in Fig. 1b, energy density is negative
everywhere for $\theta \in \left[ 0,2\pi \right] $. This is also seen from
the shape of the throat whose curvature is positive on $\theta \in \left[
0,2\pi \right] .$ Although the total exotic matter for throat of the form of
a circle of radius one is $-0.5$, in the case of (37) the total exotic
matter is $-0.48111$ in geometrical unit. This shows that a small
deformation causes the total exotic matter to be less.

\subsection{ $\protect\sigma _{0}\lessgtr 0,\Omega <0$}

As our second case we consider \ 
\begin{equation}
R_{0}\left( \theta \right) =\frac{1}{0.5\cos ^{2}\left( 3\theta \right) +1}
\end{equation}%
which admits a throat of the shape shown in Fig. 2a. The corresponding
energy density $\sigma _{0}$ is shown in Fig. 2b. As one can see the energy
density is positive wherever the curvature of the throat is negative and
vise versa. The overall energy is negative given by $\Omega =-0.39339$ unit.

\subsection{ $\protect\sigma _{0}\lessgtr 0,\Omega >0$}

For 
\begin{equation}
R_{0}\left( \theta \right) =\frac{1}{\left\vert \cos \left( \frac{5}{2}%
\theta \right) \right\vert +1}
\end{equation}%
the shape of the throat looks like a starfish as it is displayed in Fig. 3a.
The behavior of the energy density $\sigma _{0}$ is depicted in Fig. 3b. As
it is clear $\sigma _{0}$ is positive everywhere except at the neighborhood
of the corners of the throat where the curvature is positive. The total
energy, however, is positive i.e., $\Omega =0.38888$ unit.

\subsection{ $\protect\sigma _{0}>0,\Omega >0$}

For this case let's consider 
\begin{equation}
R_{0}\left( \theta \right) =\frac{1}{\sqrt{\left\vert \cos \left( \frac{5}{2}%
\theta \right) \right\vert }+1}
\end{equation}%
which is shown in Fig. 4a. In this case $\sigma _{0}$ is positive everywhere
as it is shown in Fig. 4b and the total energy is positive i.e. $\Omega
=0.40561$ unit. We should admit, however, that although $R_{0}\left( \theta
\right) $ is well defined everywhere in the range of $\theta $ i.e., $\theta
\in \lbrack 0,2\pi ]$ at five spike points in the Fig. 4a its derivative
does not exist. This feature of $R_{0}\left( \theta \right) $ causes the
energy density to be discontinuous at the same values of $\theta .$ This
means that $\sigma _{0}$ is not defined at those points although its limit
exists and is positive and finite. Fig. 4c shows that $\sigma _{0}$ is
positive around those critical points.

\subsection{Parametric ansatz for $R_{0}\left( \protect\theta \right) $}

To complete our analysis we look at the case given in Section B and
generalize the form of $R_{0}\left( \theta \right) $ as given by%
\begin{equation}
R_{0}\left( \theta \right) =\frac{1}{\epsilon \cos ^{2}\left( n\theta
\right) +1}
\end{equation}%
in which $\epsilon \in 
\mathbb{R}
^{+}$ and $n=2,3,4,...$. In Fig. 5 we plot $R_{0}\left( \theta \right) $ in
terms of different $n$ and $\epsilon =0.1.$ The total exotic matter for each
case is also calculated. We observe that increasing $n$ decreases the
magnitude of the exotic matter. When $n$ goes to infinity (i.e. an infinite
oscillation) the total energy goes to zero while at each point the energy
density shows a positive or negative fluctuation. Of course the assumption
of infinite frequency takes us away from the domain of classical physics,
probably to the quantum domain. In the latter, particle creation from a
vacuum is well-known. Herein, instead of particles we have formation of
wormholes. The ansatz (41) shows how one can go to the vacuum case $\Omega
\rightarrow 0$ with $n\rightarrow \infty .$ Yet we wish to abide by the
classical domain with $n\ll \infty .$

\section{2+1-dimensional wormhole induced by 3+1-dimensional flat spacetime}

\begin{figure}[h]
\includegraphics[width=90mm,scale=0.7]{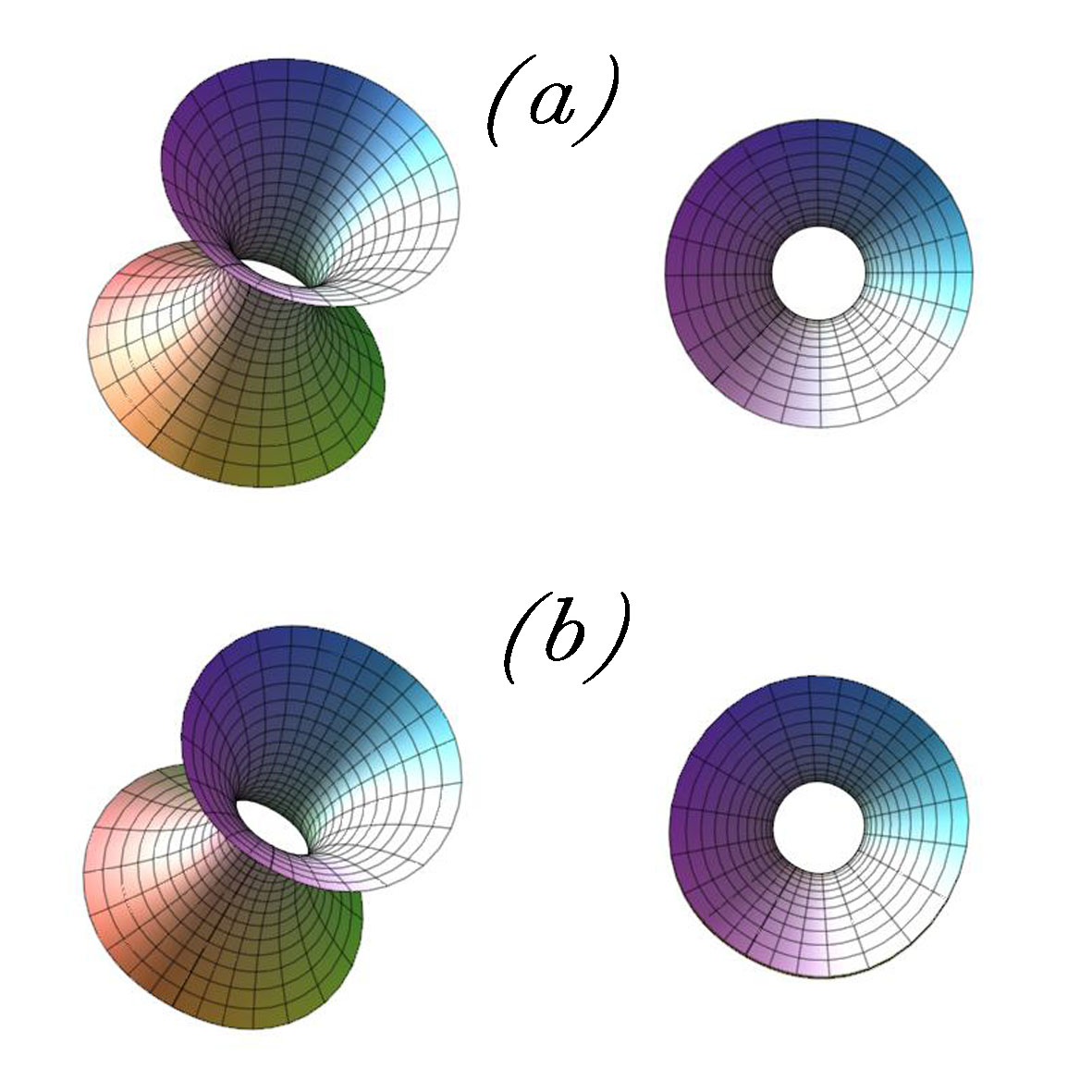}
\caption{The geometry of the throat when $R_{0}\left( \protect\theta \right)
=1$ for (a) and $R_{0}\left( \protect\theta \right) =\frac{1}{0.5\cos ^{2}%
\protect\theta +1}$ for (b) with $r_{0}=1.$ }
\end{figure}
\begin{figure}[h]
\includegraphics[width=90mm,scale=0.7]{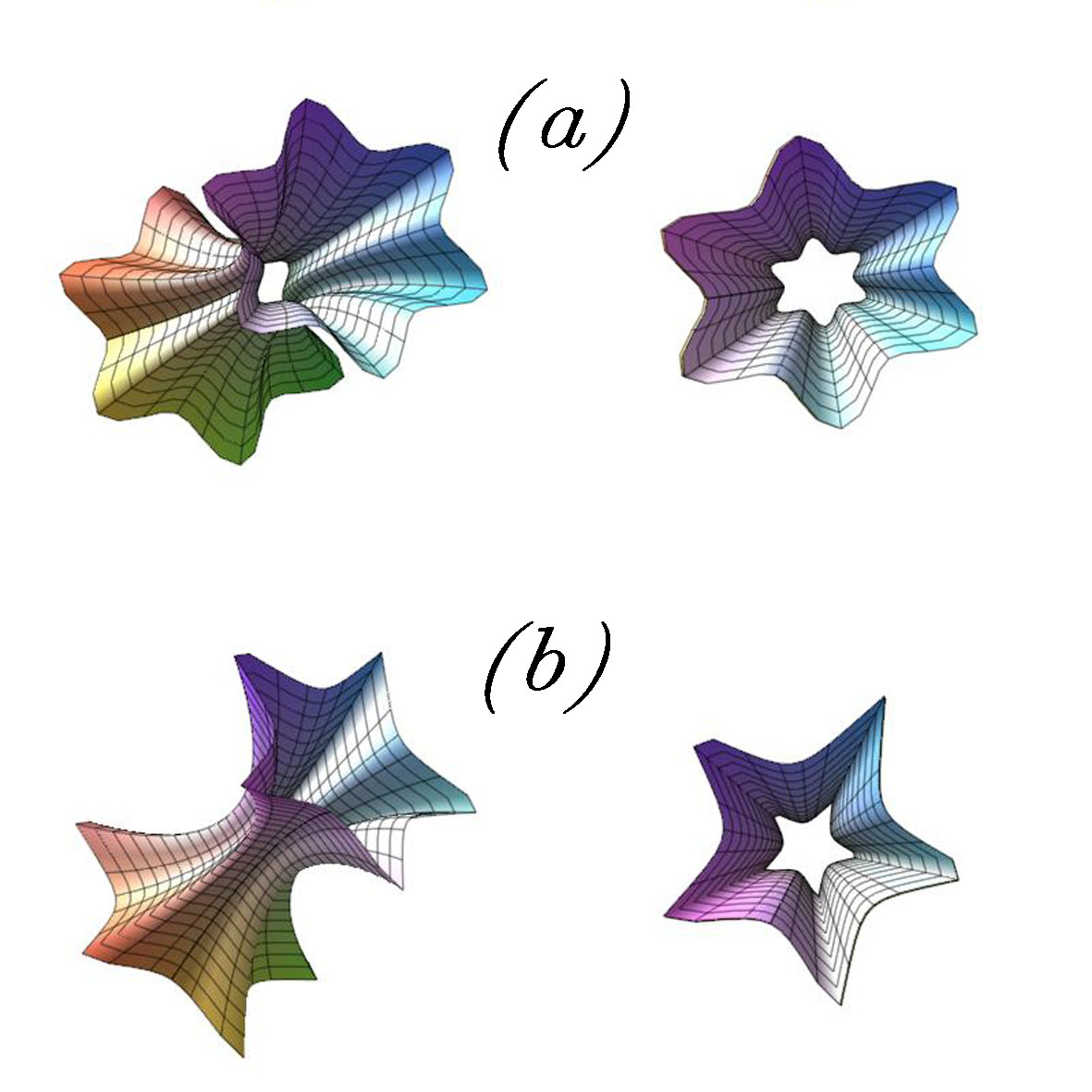}
\caption{The geometry of the throat when $R_{0}\left( \protect\theta \right)
=\frac{1}{0.5\cos ^{2}\left( 3\protect\theta \right) +1}$ for (a) and $%
R_{0}\left( \protect\theta \right) =\frac{1}{\left\vert \cos \left( \frac{5}{%
2}\protect\theta \right) \right\vert +1}$ for (b) with $r_{0}=1.$ }
\end{figure}
\begin{figure}[h]
\includegraphics[width=90mm,scale=0.7]{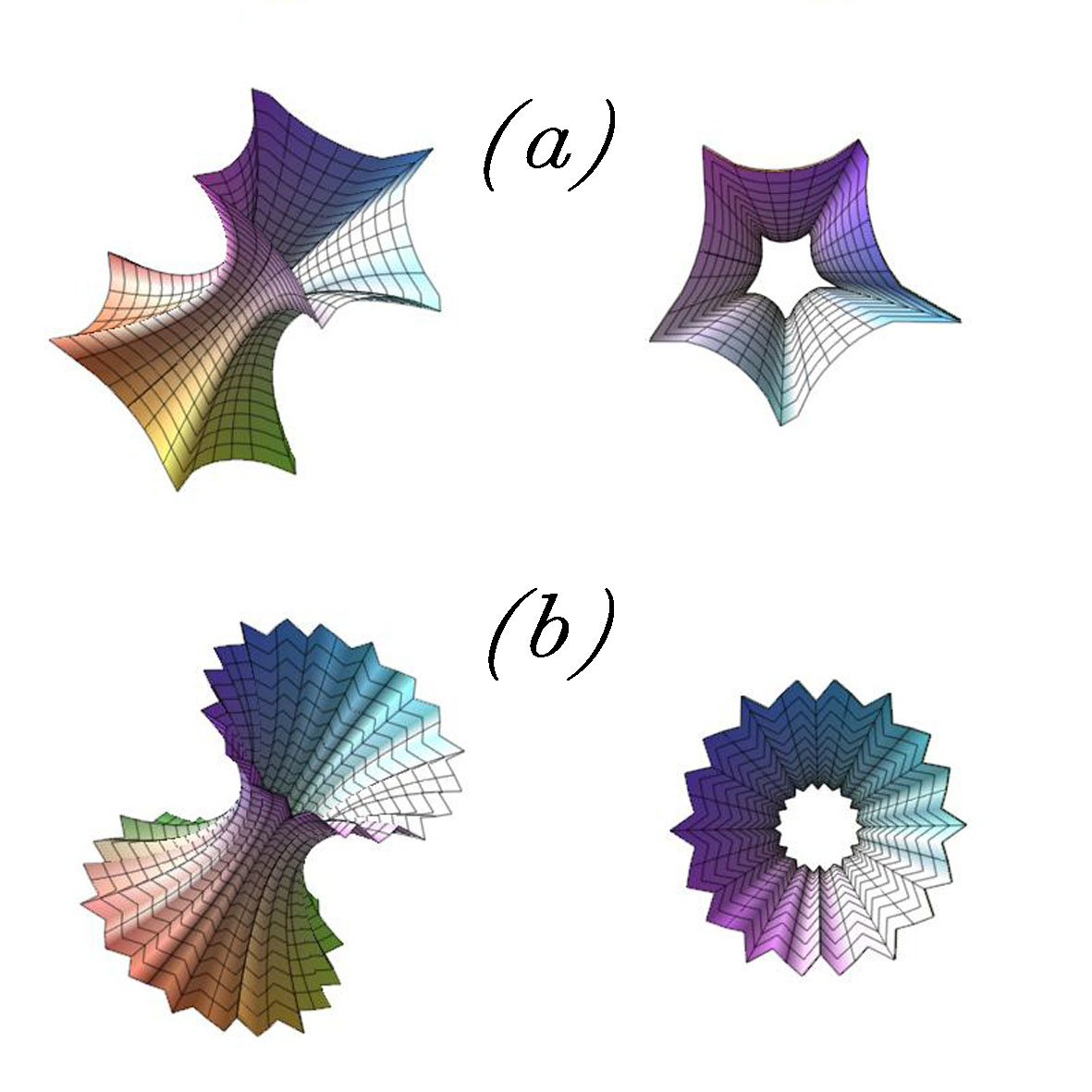}
\caption{The geometry of the throat when $R_{0}\left( \protect\theta \right)
=\frac{1}{\protect\sqrt{\left\vert \cos \left( \frac{5}{2}\protect\theta %
\right) \right\vert }+1}$ for (a) and $R_{0}\left( \protect\theta \right) =%
\frac{1}{\protect\epsilon \cos ^{2}\left( n\protect\theta \right) +1}$ for
(b) with $r_{0}=1$, $\protect\epsilon =0.1$ and $n=30.$}
\end{figure}

\begin{figure}[h]
\includegraphics[width=90mm,scale=0.7]{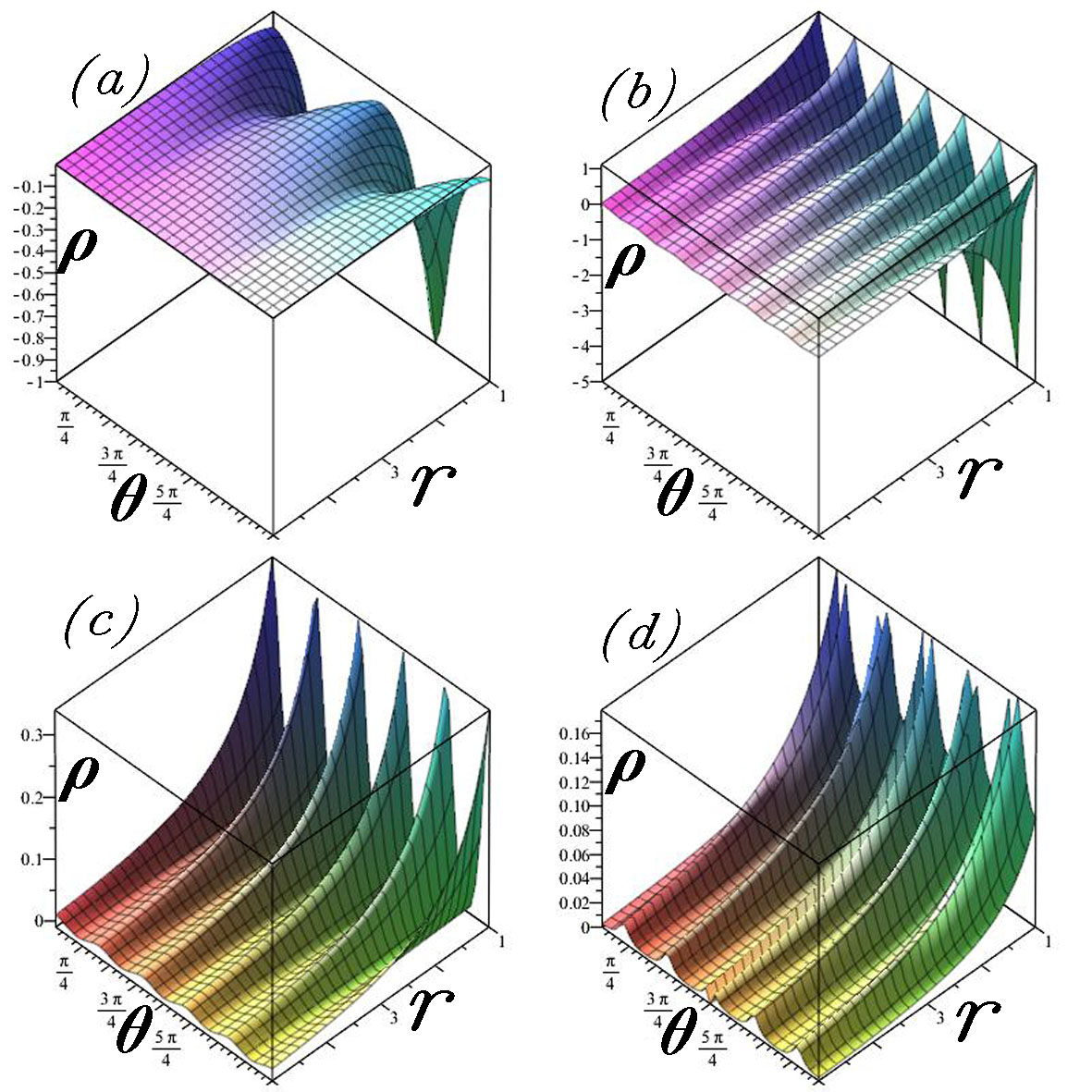}
\caption{The energy densities of the wormholes given in (37) for (a), (38)
for (b), (39) for (c) and (40) for (d). As we observe, in the cases (b) and
(c) energy density gets positive value for some interval while for (d) $%
\protect\rho >0$ everywhere. In (a) the energy density is negative
everywhere.}
\end{figure}

We consider the $3+1-$dimensional Minkowski spacetime in the cylindrical
coordinates 
\begin{equation}
ds^{2}=-dt^{2}+dr^{2}+dz^{2}+r^{2}d\theta ^{2}
\end{equation}%
with the substitution $z=\xi \left( r,\theta \right) .$ This gives the line
element%
\begin{multline}
ds^{2}=-dt^{2}+\left( 1+\xi _{r}\left( r,\theta \right) ^{2}\right) dr^{2}+
\\
\left( r^{2}+\xi _{\theta }\left( r,\theta \right) ^{2}\right) d\theta
^{2}+2\xi _{r}\left( r,\theta \right) \xi _{\theta }\left( r,\theta \right)
drd\theta .
\end{multline}%
in which $\xi _{r}\left( r,\theta \right) =\frac{\partial \xi \left(
r,\theta \right) }{\partial r}$ and $\xi _{\theta }\left( r,\theta \right) =%
\frac{\partial \xi \left( r,\theta \right) }{\partial \theta }$ where $\xi
\left( r,\theta \right) $ is a function of $r$ and $\theta $. Using this
line element, the Einstein's tensor is obtained with only one nonzero
component i.e.,%
\begin{equation}
G_{t}^{t}=-\frac{r^{3}\xi _{r}\xi _{rr}+\xi _{rr}\xi _{\theta \theta
}r^{2}-\xi _{\theta }^{2}+2\xi _{r\theta }\xi _{\theta }r-\xi _{r\theta
}^{2}r}{\left( r^{2}+\xi _{\theta }^{2}+\xi _{r}^{2}r^{2}\right) ^{2}}.
\end{equation}%
Einstein's equation ($8\pi G=1=c$) reads 
\begin{equation}
G_{\mu }^{\nu }=T_{\mu }^{\nu }
\end{equation}%
in which $T_{\mu }^{\nu }$ is the energy momentum tensor. The latter implies
that the only non-zero component of the energy-momentum tensor is $%
T_{t}^{t}=-\rho $ component and therefore 
\begin{equation}
\rho =\frac{r^{3}\xi _{r}\xi _{rr}+\xi _{rr}\xi _{\theta \theta }r^{2}-\xi
_{\theta }^{2}+2\xi _{r\theta }\xi _{\theta }r-\xi _{r\theta }^{2}r}{\left(
r^{2}+\xi _{\theta }^{2}+\xi _{r}^{2}r^{2}\right) ^{2}}.
\end{equation}%
The total energy which supports the wormhole is obtained by%
\begin{equation}
\Omega =\int_{0}^{2\pi }\int_{0}^{\infty }\rho \sqrt{-g}drd\theta .
\end{equation}%
To complete this section we recall the case of physical energy-momentum
tensor and the weak energy condition which we have discussed before. Here,
the situation is almost the same i.e., the only nonzero component of the
energy-momentum tensor is the energy density $\rho .$ Therefore in order to
say that our wormhole is physical, which means it is supported by normal
(non-exotic) matter, the conditions are $\rho \geq 0$ and $0\leq \Omega
<\infty .$

\subsection{Flare-out conditions}

To have a wormhole we observe that $z=\xi \left( r,\theta \right) $ must be
chosen aptly for a general wormhole structure. For instance, one may
consider in the first attempt $\xi \left( r,\theta \right) =\xi \left(
r\right) $ and following that we find%
\begin{equation}
ds^{2}=-dt^{2}+\left( 1+\xi ^{\prime }\left( r\right) ^{2}\right)
dr^{2}+r^{2}d\theta ^{2},
\end{equation}%
so that the form of energy density becomes%
\begin{equation}
\rho =\frac{\xi ^{\prime }\xi ^{\prime \prime }}{r\left( 1+\xi ^{\prime
2}\right) ^{2}}.
\end{equation}%
The expression (48) is comparable with the Morris-Thorne's static wormhole%
\begin{equation}
ds^{2}=-e^{2\Phi \left( r\right) }dt^{2}+\frac{1}{1-\frac{b\left( r\right) }{%
r}}dr^{2}+r^{2}d\theta ^{2},
\end{equation}%
in which $\Phi \left( r\right) $ and $b\left( r\right) $ are the red-shift
and shape functions, respectively. In the specific case (48), one finds $%
\Phi \left( r\right) =0$ and $b\left( r\right) =r\left( \frac{\xi ^{\prime
}\left( r\right) ^{2}}{1+\xi ^{\prime }\left( r\right) ^{2}}\right) .$ The
well-known flare-out condition introduced by Morris and Thorne implies that
if $r=r_{0\text{ }}$is the location of the throat, i) $b\left( r_{0}\right)
=r_{0}$ and ii) for $r>r_{0},$ $b^{\prime }\left( r\right) <\frac{b\left(
r\right) }{r}.$ In terms of the new setting, i) implies that at the throat $%
\xi ^{\prime }=\pm \infty $ and ii) states that $\xi ^{\prime }\xi ^{\prime
\prime }<0$ for $r>r_{0}.$ In addition to these conditions at the throat we
have $z=\xi \left( r_{0}\right) =0.$

Next, we introduce the location of the throat at $z=0$ and $r=R_{0}\left(
\theta \right) $ in which $R_{0}\left( \theta \right) $ is a periodic
function of $\theta .$ These mean that $z=\xi \left( R_{0},\theta \right)
=0. $ Now, for a general function for $z=\xi \left( r,\theta \right) $ we
impose the same conditions as the Morris-Thorne wormholes i.e., $\xi _{r}\xi
_{rr}<0 $ for $r>R_{0}\left( \theta \right) $ and at the location of the
throat (where $z=0$ and $r=R_{0}\left( \theta \right) $) $\xi _{r}\left(
R_{0},\theta \right) =\pm \infty .$

\subsection{An Illustrative Example}

Here we present an explicit example. Let's consider 
\begin{equation}
\xi =\pm 2r_{0}\sqrt{\left( \frac{r}{R_{0}\left( \theta \right) }-1\right) }
\end{equation}%
in which the first condition i.e., $\xi =0$ at the location of the throat $%
r=R_{0}\left( \theta \right) $ is fulfilled. Next, the expression, 
\begin{equation}
\xi _{r}\xi _{rr}=-\frac{1}{2}\frac{r_{0}^{2}}{\left( \frac{r}{R_{0}}%
-1\right) ^{2}R_{0}^{3}}<0
\end{equation}%
imposes $R_{0}\left( \theta \right) >0$ on the entire domain of $\theta $
i.e. $\theta \in \left[ 0,2\pi \right] .$ We note also that $R_{0}\left(
\theta \right) $ must be a periodic function of $\theta $ to make $%
r=r_{0}R_{0}\left( \theta \right) $ a closed loop, which is going to be our
throat. The forms of 
\begin{equation}
\xi _{r}=\frac{\pm r_{0}}{R_{0}\sqrt{\frac{r}{R_{0}}-1}}
\end{equation}%
and%
\begin{equation}
\xi _{rr}=-\frac{\pm r_{0}}{2R^{2}\left( \frac{r}{R_{0}}-1\right) ^{3/2}}
\end{equation}%
suggest that at the throat $\xi _{r}\rightarrow \pm \infty $ and also $\xi
_{rr}\rightarrow \mp \infty $ as expected. The form of energy density in
terms of $R$, however, becomes%
\begin{equation}
\rho =\frac{R_{0}^{3}r_{0}^{2}\left( R_{0}^{\prime \prime
}R_{0}-R_{0}^{2}-2R_{0}^{\prime 2}\right) }{2r\left( r_{0}^{2}R_{0}^{\prime
2}+R_{0}^{2}\left( r_{0}^{2}+rR_{0}-r_{0}R_{0}^{2}\right) \right) ^{2}}.
\end{equation}%
The latter implies that any periodic function of $R_{0}\left( \theta \right) 
$ which satisfies $R_{0}^{\prime \prime }R_{0}-R_{0}^{2}-2R_{0}^{\prime 2}>0$
can represent a traversable wormhole with positive energy. In the case of $%
R_{0}=r_{0}$ or $\xi =\pm 2r_{0}\sqrt{\frac{r}{r_{0}}-1}$, the wormhole is
shown in Fig. 6 whose energy density is given by%
\begin{equation}
\rho =-\frac{r_{0}}{2r^{3}}.
\end{equation}

In Figs. 6-8 we plot the wormholes with $R_{0}\left( \theta \right) $ given
in (37), (38), (39), (40) and (41), respectively. Also in Fig. 9, the energy
density $\rho $ corresponds to the individual cases of (a), (b), (c) and (d)
for (37), (38), (39) and (40), respectively, which are given in terms of $r$
and $\theta $ with $r_{0}=1.$We see, for instance, that in Fig. 9d the
energy is positive everywhere.

Regarding Fig. 9d we must admit again that for the same reason as in the
TSW, $R_{0}\left( \theta \right) $ does not admit derivative at the spike
points, therefore the energy density $\rho $ is not defined at the same
points. On the other hand its limits exist at these critical points and are
both finite and positive. This can be seen from (55) whose both numerator
and denominator become infinity near the critical points while the ratio
which is the limit of $\rho ,$ remains finite. This is similar to the
behavior of a function like $\frac{\sin \left( x\right) }{x}$ in the
neighborhood of $x=0.$

\section{Conclusion}

First of all let us admit that traversable wormholes in $2+1-$dimensions
were considered much earlier, namely in 1990s \cite{Mann}, and became
fashionable recently \cite{Revised}. Our principal aim in this study is to
establish a traversable wormhole with normal (i.e. non-exotic) matter in $%
2+1-$dimensions. For the TSWs the strategy is to assume a closed angular
path of the form $r=R\left( t,\theta \right) $, where for $R\left( t,\theta
\right) =a\left( t\right) $, we recover the circular throat topology. Since
this leads to exotic matter it is not attractive by our assessment. Let us
note that throughout our study we use the coordinate time instead of the
proper time. Our analysis shows that any concave-shaped $R_{0}\left( \theta
\right) =R\left( 0,\theta \right) $ around the origin undergoes the same
fate of exotic matter. However, a convex-shaped $R_{0}\left( \theta \right) $
seems promising in obtaining a normal matter. This is shown by explicit
ansatzes whose plots suggest starfish-shaped closed curves for the throat of
a $2+1-$dimensional wormhole. Locally, for specific angular range it may
yield negative energy, but in total the energy accumulates on the positive
side. This result supports our previous finding that for the non-spherical
spacetimes, i.e. the ZV-metrics, the throats can be non-spherical and in
turn one may obtain a TSW in Einstein's theory with a positive total energy 
\cite{Habib2}. The conclusion drawn herein for $2+1-$dimensions therefore
can be generalized to higher dimensions. A similar construction method has
been employed for the general wormholes. We have treated the $2+1-$%
dimensional wormhole as a brane in $3+1-$dimensional Minkowski space and
show the possibility of physical traversable wormhole. It turns out,
however, that the rabbit emerges from only very special hats, not from all
hats. Extension of our work to $3+1-$dimensional wormholes is under
construction.

\bigskip

\end{document}